\documentclass[universe,communication,accept,pdftex,oneauthor]{Definitions/mdpi} 

\firstpage{1} 
\pubvolume{1}
\issuenum{1}
\articlenumber{0}
\pubyear{2022}
\copyrightyear{2022}
\externaleditor{{Academic Editor: Lorenzo Iorio} 
 } 
\datereceived{26 September 2022} 
\dateaccepted{26 October 2022} 
\datepublished{} 
\hreflink{https://doi.org/} 
\pdfoutput=1

\usepackage{amsmath}

\newcommand{\eq}[1]{\begin{linenomath}
\begin{equation}
\begin{aligned}
#1
\end{aligned}
\end{equation}
\end{linenomath}}

\newcommand{\pa}{\partial}

\newcommand{\dtfrac}[2]{\dfrac{\delta #1}{\delta #2}}
\newcommand{\expval}[1]{\langle{#1}{\rangle}}

\newcommand{\mH}{\mathcal{H}}
\newcommand{\mV}{\mathcal{V}}
\newcommand{\mL}{\mathcal{L}}

\Title{A {Wheeler--DeWitt} Equation with Time} 

\TitleCitation{A Wheeler--DeWitt Equation with Time}

\Author{Marcello Rotondo \orcidA}

\AuthorNames{Marcello Rotondo}

\AuthorCitation{Rotondo, M.}

\address[1]{%
 {Independent Researcher; Torino 10142, Italy} 
 marcello@gravity.phys.nagoya-u.ac.jp}

\abstract{The equation for canonical gravity produced by Wheeler and {DeWitt} in the late 1960s still presents difficulties both in terms of its mathematical solution and {its} physical interpretation. One~of these issues is, notoriously, the absence of an explicit time. In this short note, we suggest one simple and straightforward way to avoid this {occurrence}. We go back to the classical equation that inspired Wheeler and {DeWitt} {(namely, the Hamilton--Jacobi--Einstein equation)} and make explicit{, before quantization,} the presence of a known, classically meaningful notion of time. {We do this by} allowing Hamilton's principal function to be {explicitly} dependent on this time locally. This choice results in a {Wheeler--DeWitt} equation with time. A working solution for the de Sitter minisuperspace is shown.}

\begin{document}

\section{Introduction}

One traditional avenue to the quantization of gravity is the geometrodynamical one, represented by the infamous {Wheeler--DeWitt} ({WDW}) equation \cite{DEW(1967),WHE(1968)}. The equation is expected to describe the quantum evolution of the spatial components of the metric tensor of General Relativity (GR), but its solution and interpretation are long-standing problems~\cite{ROV(2004)}. In particular, a problem with time occurs when we try to {interpret} the {WDW} equation as a Schr\"odinger-type equation for gravity, {because} the state it describes appears to be~stationary.

{To begin with,} the absence of time from the {WDW} equation is a consequence of the fact that the first-class Hamiltonian constraint of GR, of which the {WDW} equation intends to be the quantization, specifically enforces time diffeomorphism. In other words, it ensures its dynamical laws are valid {independently} of our choice of time coordinate. When we consider the so-called Hamilton--Jacobi--Einstein (HJE) equation developed by Peres \cite{PER(1962)}, which expresses the constraint on the 00 component of the Einstein field equations in the Hamilton--Jacobi formalism, it is clear that in the classical case, time is absent where it should appear, even though the theory is classical. We know, however, that the HJE equation does not describe a timeless geometry. The reason that the HJE equation is not problematic can be traced back to the fact that a notion of time exists for the evolution of the spatial geometry, as long as the classical notion of trajectory in superspace holds. In that sense, it appears that the actual problem with time is not that it is absent in the {Schr\"odinger}-type equation itself, but that we cannot introduce it as we do in the classical case, since space no longer evolves along classical trajectories. {It is not clear what becomes of this time beyond the semi-classical level, when gravity does not act as a stage for matter fields, but rather partakes in the quantum dance.}

{The absence of an external time in the description of GR, which is inherited by its quantization, is sometimes referred to as the ``frozen formalism problem'', and constitutes only one among other difficulties in the definition of time in classical and quantum physics~\cite{AND(2017)}. In the present work, we will address only this particular aspect of the problem, ignoring its relations to others (a strictly related one being the definition of time-evolving observables for quantum gravity). Two classic reviews on this subject are Isham's \cite{ISH(1993)} and Kucha\v{r}'s~\cite{KUC(1992)}. We invite the reader to read these reviews for detailed references, and to gain a general idea of the large extent of variable approaches. The study of this aspect of the problem of time has certainly evolved significantly since the time of these reviews, with some issues of each approach being successfully addressed, but it ultimately remains open. Faced by the menace represented by the loss of a useful notion of time, three alternative reactions have been adopted by researchers: \emph{{flight}}, \emph{{fight}}, or {\emph{freeze}} (corresponding to Isham's \emph{tempus ante quantum}, \emph{tempus post quantum}, and \emph{tempus nihil est}).
}

\begin{enumerate}

\item { \emph{Flight}: Time is recognized as a fundamental element of our description of physical phenomena, and attempts are made to define it \emph{before} quantization, as a functional of the canonical variables. This is a conservative approach that tries to obtain an ``external'' notion of time as that appearing in Schr\"odinger's equation.}

\item
{\emph{Fight}: Time is recognized as a fundamental element of our description of physical phenomena, but it is retrieved only \emph{after} the quantization. This type of approach fights against the interpretative problems presented by the quantum theory to obtain a novel definition of time.}

\item
{
\emph{Freeze}: Timelessness is accepted, time is forsaken as a fundamental notion for the description of quantum gravity, and attempts are made to provide a complete quantum theory otherwise.}

\end{enumerate}

{The present work adopts a definition of time resulting from the semi-classical approximation of the WDW equation, an approach falling within the second category above. However, we do not limit to the semi-classical regime the definition of time identified by the semi-classical approximation.} The point of the present note is {in fact} to suggest that {the ``frozen formalism'' could} be avoided by {retaining the use of} a classical notion of time suggested by the {semi-classical} approximation of the theory, even though quantum space does not evolve along classical trajectories. Therefore, our proposal belongs to the first category, in that it carries over to the quantum regime the definition of time justified by the semi-classical approximation. 

{The definition of time that we adopt as a starting point is that naturally resulting from the semi-classical approximation of canonical quantum gravity obtained by expanding the total wave functional in inverse squares of the Planck mass \cite{KIE(1991), KIE(1994),KIE(2007)}. Classical GR and the Schr\"odinger equation for non-gravitational fields straightforwardly recovered this approximation. Time, for the matter fields, is a multi-fingered (i.e., space dependent) functional generated by the classical evolution of background geometry along its trajectory in superspace. The operation has formal analogies with the {Wentzel--Kramers--Brillouin (WKB)} approximations of quantum {mechanics}, and the Born--Oppenheimer approximation from molecular physics \cite{BER(1996),KAM(2017),KAM(2019)}. In recent years, special attention was drawn to the problem of re-establishing the unitarity of time evolution in this approach (see, for example, \cite{KIE(2018),CHA(2019), CHA(2020), GIO(2022)}). For a recent work by this author which is related to the present one, see \cite{ROT(2020)}.}

{One key observation that motivated this approach was that} this multi-fingered (or ``WKB'') functional time can be applied to the HJE equation itself, and we can make the presence of that classical time explicit in the HJE equation, which is the timeless form. This alone was the starting point of Wheeler and DeWitt, as the following memoir by one of the authors recalls.

\begin{quotation}
One day in 1965, John Wheeler had a two-hour stopover between flights at the Raleigh--Durham airport in North Carolina. He called Bryce DeWitt, then at the University of North Carolina in Chapel Hill, proposing to meet at the airport during the wait. Bryce showed up with the Hamilton--Jacobi equation of general relativity, published by Asher Peres not long before [...] Bryce mumbled the idea of repeating what Schr\"oedinger did for the hydrogen atom: obtaining a wave equation by replacing the square of derivatives with (i times) a second derivative---a manner for undoing the optical approximation. [...] Wheeler was tremendously excited (he was often enthusiastic) and declared on the spot that the equation for quantum gravity had been found. \cite{ROV(2015),DEW(2011)}
\end{quotation}

What if DeWitt had presented Wheeler with the HJE equation \textit{together with} the notion of multi-fingered time? The resulting equation does present the functional time variable just as time appears in the {Schr\"odinger} equation, and preserves the correct classical limit for gravity.

In Section \ref{1}, we briefly review the definition of this multi-fingered time. In Section \ref{2}, we {rewrite the HJE equation, allowing an explicit local dependence of Hamilton's principal function on that time,} and write the associated {WDW} equation with time. Finally, in Section~\ref{3}, we discuss a simple realization in the de Sitter minisuperspace that is of special interest to quantum cosmology.


\section{Classical Time Evolution from the Hamiltonian Constraint} \label{1}

{For a straightforward introduction to the emergence of time in the semi-classical approximation of canonical quantum gravity, we refer the reader to the mentioned work by Isham \cite{ISH(1993)}, {Section 5.4} 
and reference{s} therein. Here, we follow, with some variation, the notation adopted by Kiefer \cite{KIE(1994)}.}
Consider a generic spacetime with {line~element}

\eq{
	ds^2 = - {N^2} dt^2 + N_i dt dx^i + h_{ij} dx^i dx^j \, . \label{eq:ds}
}

Here, $h_{ij} = g_{\mu\nu},  \mu,\nu \in \{ 1,2,3\}$ is the spatial metric, with Latin indices $i,j \in( 1,2,3)$, $N_i = g_{0i}$ is the shift function, and $N = (- g^{00})^{-1/2}$ is the lapse function. In the Hamiltonian formalism of GR, time parametrization invariance {is} enforced by a first-class Hamiltonian constraint {(i.e., a constraint imposed on the Hamiltonian only after the equations of motion are satisfied).}

\eq{
	\int d^3x \left[ (2M)^{-1}  G_{AB} \pi^A \pi^B + \mV(h_A) + \mH_\phi(\phi;h_A) \right] = 0 \, . \label{eq:HC}
}

(This constraint can be obtained from the variation of the super-Hamiltonian of GR with respect to the lapse function $N$ in the {Arnowitt--Deser--Misner (ADM)} formalism \cite{ADM(2008)}. Here, we intend to recall only the elements strictly necessary to follow our discussion.)

In {Equation} \eqref{eq:HC}, the capital indices, {$A,B = \{ij\}$} represent pairs of {Latin indices}, and $G_{AB}$ is the {DeWitt} metric

\eq{
	G_{AB} \equiv G_{ijkl} = \frac{1}{2\sqrt{h}} \left( h_{ik}h_{jl} + h_{il}h_{jk} - h_{ij}h_{kl}\right) \, ,
}
{{underlying} superspace, i.e., the space of spatial metrics up to differeomorphism~invariance.}

The physical scale (we have set $\hbar = c = 1$) of quantum gravity is set by the ``geometrodynamical mass'' $M$, which is proportional to the square of the Planck mass $m_{P}$

\eq{
	M = ( m_P / 2 )^2 \, , \quad m_P = \left( 8 \pi G \right)^{-1/2} \, .
}

The geometrodynamical potential density $\mV$ is

\eq{
	\mV = 2 M \sqrt{h} ( 2 \Lambda - {}^{(3)}R) \, ,
}
where $h$ and ${}^{(3)}R$ are the determinant and the Ricci scalar of the spatial metric, respectively. The Hamiltonian density operator $\mH_\phi$ is taken to describe bosonic matter.

The {WDW} equation results from an attempt to quantize the Hamiltonian constraint~\eqref{eq:HC} \textls[-25]{straightforwardly, applying it to the ``wave functional of the universe'', $\Psi[h_A,\phi]$, \mbox{thus~obtaining}}

\eq{
	\int d^3x \left[ (2M)^{-1} G_{AB}\pa^A \pa^B + \mV(h_A) + \mH_\phi(h_A, \phi) \right] \Psi[h_A,\phi] = 0 \, , \label{eq:WDW}
}
where all variables are promoted to the respective operators. For the sake of simplicity, in this section, we have adopted the trivial ordering, and the symbols $\pa_A$ are used to indicate functional derivatives with respect to the metric component indicated by the double index.

The evolution of bosonic quantum fields in classical curved spacetime is obtained by making the ansatz

\eq{
	\Psi[h_A, \phi] = \chi[h_A] \psi[\phi;h_A]
}

for the total wave functional, and considering a WKB-like expansion in inverse powers of $M$ \cite{KIE(1994), KIE(2007)}. In doing so, one aims at wave functionals $\chi$ and $\psi$ that describe the ``heavy'' (i.e., the spatial metric components) and ``light'' degrees of freedom (i.e., matter), respectively. Notice that $\psi$ depends on the geometry only parametrically, which is indicated by the use of the semicolon. The method consists of substituting the expansion

\eq{
	\Psi[h_A, \phi] = \exp \left( i \sum_{n=0}^\infty M^{1-n} S_n[h_A, \phi] \right) \label{eq:S}
}
in the {WDW} equation, and equating contributions to equal powers of $M$.

To order $M^2$, one obtains $S_0 = S_0[h_A]$: the leading contribution is purely geometrodynamical. To order $M^1$, one obtains the vacuum HJE equation

\eq{
	\int d^3x \left[ (2M)^{-1} G_{AB} \pa^A S_G \pa^B S_G + \mV \right]= 0 \, , \label{eq:HJE}
}

$S_G = M S_0$ being the leading contribution to the phase of the wave functional \eqref{eq:S}. The HJE {Equation} \eqref{eq:HJE} appears to be timeless due to the vanishing of the RHS. However, the time evolution of space can still be obtained from the Hamiltonian constraint \eqref{eq:HC} by expressing the canonical momenta (defined by the Lagrangian as $\pi^A = \pa \mL / \pa \dot{h}_A$, and identified with $\pi^A = \pa^A S_G$ in the Hamilton--Jacobi formalism) in terms of the geometrodynamical velocities. From the Hamiltonian equations of motion in the ADM formalism, these are given by
\eq{
	\dot{h}_{A} = N \pi_A + 2 N_{(A)} \, , \label{eq:ht}
}

$N_{(A)}$ being a shorthand for $N_{(i;j)}$. The fact that time evolution is retrieved by such substitution, as the Hamiltonian is constrained, is an important point.

At this point, define $\psi_1[\phi; h_A] = \exp \left( i S_1[\phi, h_A] \right)$ and require conservation of the current associated with $\chi$. Then, to order $M^0$, one gets the following functional equation for~matter
\eq{
	\int d^3x \left[ i M^{-1} G_{AB} \pa^A S_G \pa^B  - \mH_\phi \right] \psi_1[\phi; h_A] = 0 \label{eq:SE} \, .
}

By using as time the local ``multi-fingered'' time $\tau = \tau(x)$ of the parametrization generated by the classical momenta along the classical trajectory in superspace

\eq{
	 G_{AB} \pi^A \pa^B\Big|_{x} \tau(y) = \delta(y-x) \, ,
}

Equation \eqref{eq:SE} gives the Schr\"odinger-type functional equation

\eq{
	\int d^3x \left[ i \overset{\circ}{\psi}_1[\phi; h_A]  - \mH_\phi \psi_1[\phi; h_A] \right] = 0 \label{eq:SE2} \, ,
}
where we employ a circle over the variable to indicate the functional derivative with respect to $\tau$
\eq{
	\dtfrac{}{\tau} = M^{-1} G_{AB} \pi^A \pa^B \label{eq:WKBT1} \, .
}

Notice that in normal coordinates ($N=1$, $N_{0i} = 0$), the {Equation} \eqref{eq:SE2} reduces to the functional {Schr\"odinger} equation for bosonic fields. In this case, the passage to the partial derivative with respect to time is granted by the fact that the matter wave functional $\psi_1[\phi; h_A]$ depends explicitly on time only through the background metric, which appears as a set of local parameters, and employing the resulting relation between momenta and velocities \eqref{eq:ht}, \eqref{eq:WKBT1} reduces to an application of the chain rule.


\section{The Time Evolution of Quantum Space} \label{2}

The functional derivative with respect to $\tau$ of a metric component, intended as \mbox{the~functional}
\eq{
	h_A(x) = \int d^3y \, h_A(y) \, \delta(y-x) \, ,
}
gives the relation between velocities and momenta,
\eq{
	\overset{\circ}{h}_A = M^{-1} G_{AB}\pi^A \, . \label{eq:hdt}
}

Putting the relation \eqref{eq:hdt} back into the Hamiltonian equation gives the equations of motion with respect to $\tau$. In other words, we can use $\tau$ not only to describe the evolution of quantum fields with respect to the background metric, but also to describe the evolution of the background metric itself. It is a favorable parametrization in that it makes the form of the equations of motion simpler and not explicitly dependent on the coordinate choice \eqref{eq:ds}. Incidentally, notice that, in Peres' HJE equation, the spatial metric components are defined as functions of space alone. How they become a function of time seems to be a problem that is not addressed in the literature. In the present treatment, they become functions of time precisely by {\em{defining}} their dependence on time according to \eqref{eq:hdt}.

The main objection in extending the use of multi-fingered time to the quantum evolution is that classical trajectories are lost in that regime. While this is true, we still know what the classical trajectory is, and we can use the ``natural'' parametrization (i.e., {the parametrization} in which all equations of motion appear simple as \eqref{eq:hdt}) along it to describe evolution along non-classical trajectories between wave fronts of constant {multi-fingered}~time.

Working with the vacuum model, we can make multi-fingered time explicit in the HJE {Equation} \eqref{eq:HJE} by rewriting it as

\eq{
	\int d^3x \left[ \pa_\tau{S_G} + (2M)^{-1} G_{AB} \pa^A S_G \pa^B S_G + \mV \right]= 0 \, , \label{eq:HJE2}
}
and requiring that
\eq{
	\int d^3x \pa_\tau{S_G} = 0 \, . \label{eq:gc}
}

Notice that the derivative with respect to $\tau$ is partial. As we previously observed, WKB time \eqref{eq:WKBT1} only takes into account the dependency on $\tau$ through the geometrodynamical degrees of freedom, but $S_G$ could, in principle, also be explicitly dependent on it. {What we are doing is simply allowing for this possibility.} Adding and subtracting the $\tau$-derivative of $S_G$, and substituting the velocities \eqref{eq:hdt}, integration of the HJE {Equation} \eqref{eq:HJE2} tells us that $S_G$ is indeed Hamilton's principal function, defined as a functional integral of the Lagrangian.

Notwithstanding the fact that {the condition} \eqref{eq:gc} ensures that the Hamiltonian constraint still holds, it allows for the action to remain dependent explicitly on $\tau$ {locally}. We may then try to quantize the HJE {Equation} \eqref{eq:HJE2} à la Schr\"odinger, and require the global condition \eqref{eq:gc} to be true only in the classical limit. What we obtain is a {WDW} equation with time
\eq{
	\int d^3x \left[ - i \frac{\pa \Psi}{ \pa \tau} + \left( (2M)^{-1} G_{AB} \pa^A \pa^B + \mV \right) \Psi \right]= 0 \, . \label{eq:WDW2}
}

Both the introduction of the coordinate independent functional time derivative \label{eq:WKBT1} and the (classically redundant) condition \eqref{eq:gc} are necessary to obtain {Equation} \eqref{eq:WDW2}. The second condition is somehow reminiscent of another approach \cite{NIK(2003)}, where time is recovered by weakening the classical Hamiltonian constraint, required to hold only on average in the quantum regime. In that work, the problem of which time to use to describe the evolution is not addressed, and the condition on time dependence is stricter than ours (see (6) in the referenced paper), resulting in a wave function whose phase does not depend on time even~locally.

\section{The de Sitter Minisuperspace} \label{3}

The results of the previous section, i.e., the time evolution {Equation \eqref{eq:WDW2}} combined with the global condition \eqref{eq:gc}, are only formal. Equation {\eqref{eq:WDW2}} still presents the same issues {as the ``timeless'' WDW Equation} \eqref{eq:WDW}. {Besides that}, using a spatial-dependent $\tau$ to parametrize the spatial geometry is more easily said than done. In the following, we will consider, by way of example, the solution for the {spatially} flat de Sitter universe, described by the line element
\eq{
	ds^2 = - dt^2 + a(t)^2 \delta^{ij} dx_i dx_j \, .
}

Here, $a(t)$ is the scale factor. {We will set $M = 1 / 12$.}

As a geometrodynamical variable, rather than the scale factor itself. It will be convenient to adopt
\eq{
	q = \frac{(2a)^{\frac{3}{2}}}{3} \left(\int d^3x\right)^{\frac{1}{2}} \, \label{eq:q} ,
}
{which is proportional to the square root of the co-moving spatial volume considered}.

The Ricci scalar of the spatial metric {vanishes} in this model, and the Hilbert--Einstein action Lagrangian, reads simply
\eq{
	L = - \frac{1}{2} \left( \dot{q}^2 + \omega^2 q^2 \right) \, .
}

{Here,} we have defined the constant $\omega = \sqrt{3\Lambda/4}$. The canonical momentum is $ \pi_q = - \dot{q} $, and the Hamiltonian is
\eq{
	H = {-}\frac{1}{2} \pi_q^2 + \frac{1}{2} \omega^2 \, q^2 
}

The Hamiltonian constraint then gives the Friedmann equation
\eq{
	\left( \frac{\dot q}{q} \right)^2 = \omega^2 \, .
}

This allows us to obtain the classical time evolution of spatial volume
\eq{
	q(t) =  {q(0)}\exp\left( \omega t\right) \, . \label{eq:at}
}

The HJE {Equation} \eqref{eq:HJE} reduces to
\eq{
	- \frac{1}{2} \left( \frac{\pa S}{\pa q} \right)^2 + \frac{1}{2} \omega^2 \, q^2 = 0 \, . \label{eq:HJE_ms}
}

Classically, Hamilton's principal function depends on time only implicitly, and is of the form
\eq{
	S = \mp \frac{\omega}{2} \left(q^2 - q_i^2\right)  \, . \label{eq:S0}
}

Using {Equation} \eqref{eq:at}, one can check that the (one-fingered) time associated with this action indeed coincides with forward coordinate time when we choose the negative sign.

Moving {onto} the quantization, notice that with our choice of variable, the DeWitt metric, that in terms of the scale factor reads $G_{aa} = - 2a$, is now
\eq{
	G_{qq} = - (3q)^{\frac{2}{3}} \, .
}

This simplifies the measure
\eq{
	\sqrt{-G_{(a)}} \, da \to dq
}
and, adopting the Laplace--Beltrami operator ordering for the kinetic operator, we have

\eq{
	\sqrt{- G_{(a)}}^{-1}  \pa_a \left( \sqrt{- G_{(a)}} G^{aa} \pa_a \right) \to  \pa_q^2 \, .
}

Then, in this minisuperspace, the {WDW} {Equation} \eqref{eq:WDW2} reads

\eq{
	i \dot \Psi = \frac{1}{2} \pa_q^2 \Psi + \frac{1}{2} \omega^2 q^2 \Psi \, , \label{eq:iho}
}
which is essentially the one-dimensional {Schr\"odinger} equation for the so-called inverted harmonic oscillator. See \cite{SUB(2021)} for a recent review, and \cite{GUT(1985)} for an application to the study of a scalar field in slow-roll inflation. The {difference} in our case is that the variable is constrained to the positive axis only, and time appears with the opposite sign. {As in \cite{GUT(1985)},} we {require} the {Gaussian} ansatz
\eq{
	\psi(q,t) = A(t) \exp \left( - B(t) q^2 \right) \, .
}

{The equality in \eqref{eq:iho} of the terms of null and second order in $q$ imposes}
\eq{
	- i \dot{A} = A \, B \label{eq:A}
}
and
\eq{
	- i \dot{B} = \frac{1}{2} \omega^2 + 2 B^2  \, . \label{eq:B}
}

From {the evolution Equation} \eqref{eq:B}, we have

\eq{
	B(t) = \frac{\omega}{2} \tan \left( \phi + i \omega t \right) \, .
}

Here $\phi$ is the real part of the constant of integration. The imaginary part of the constant is {merged into} the choice of initial time, {instead,} so that the width of the state is minimized at $t=0$. {The value of $\phi$ determines the greater ($<$$\pi$/4) or lesser ($>$$\pi$/4) peaking of the distribution in $q$, rather than the conjugate momentum at time $t=0$.}

Substituting $B(t)$ in {Equation} \eqref{eq:A}, for a normalized solution, we obtain
{
\eq{
	A(t) = \left(\frac{2}{\pi} \omega \sin(2\phi) \right)^{1/4} \left( \cos(\phi + i \omega t ) \right)^{-\frac{1}{2}} \, .
}
}\vspace{-3pt}

The expectation value {for} $q$ is
{
\eq{
	\expval{q} &= (2\pi \omega \sin(2\phi) )^{-\frac{1}{2}} \frac{\cos(2\phi) + \cosh(2 \omega t)}{\vert\cos(\phi+i \omega t)\vert} \, .
}
\vspace{-3pt}}

\begin{figure}[H]
\includegraphics[scale=0.175]{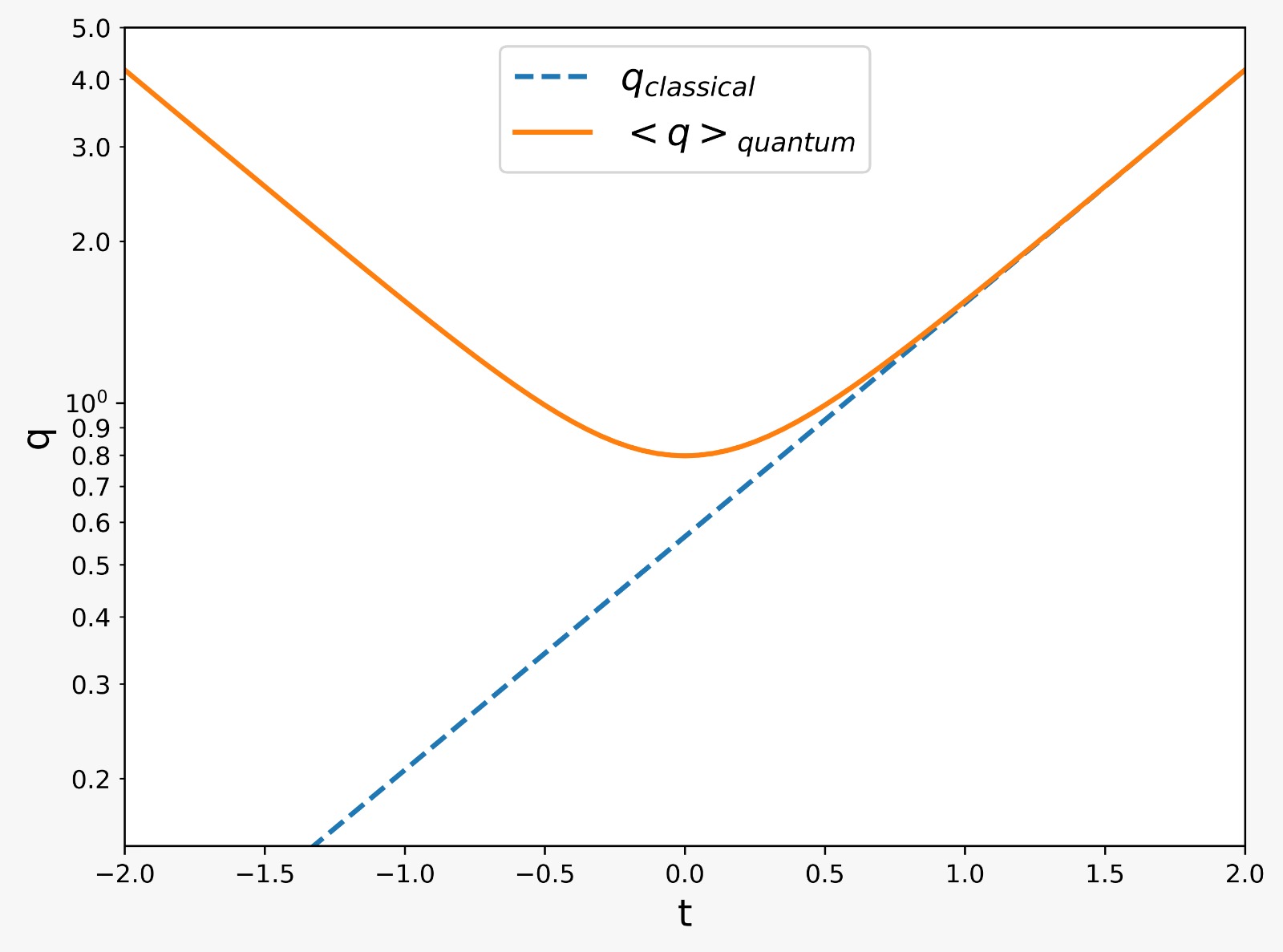}
\caption{Time evolution of the expectation value of the scale variable \emph{q} for $\phi = \pi / 4$. We set $\omega=1$.}
\centering
\label{fig:q}
\end{figure}

{At late times, we correctly recover the classical inflationary expansion $\expval{q} \propto \exp(\omega t)$ of Equation \eqref{eq:at}}. {In} this limit{,} the phase of $\psi$ approximates the classical action \eqref{eq:S0} (up to a global phase{, which fixes} the initial value) and loses its explicit dependence on time, thus {reducing to the classical action} \eqref{eq:gc}. {On the other hand, when we approach the time $t=0$, where the state is maximally contracted for the given value of $\phi$, the expectation value of the scale variable deviates from the classical one, which is directed at asymptotic convergence to zero: we observe instead the state of the de Sitter universe bouncing back from an earlier phase of contraction (see Figure \ref{fig:q}).}

\section{Conclusions}

In this short note, we have proposed to extend {to the quantum regime} the use of {the} multi-fingered time originating from the {semi-classical WKB} approximation of geometrodynamics. We have {done so by rewriting} the classical HJE {equation} for vacuum space to include {an} explicit time dependency {on} Hamilton's principal function, {and requiring this dependency to vanish globally}. {The quantization} provides a {WDW} equation that describes the evolution of the state with respect to classical multi-fingered time. {Quantum matter fields can be included by appropriately augmenting the Hamiltonian operator. The classical limit will still be granted.}

{The main purpose of this work was to provide a formal WDW equation, \eqref{eq:WDW2}, with a clear and working notion of time. This result, however, does not} help {with} the original mathematical difficulties of the {WDW} equation, such as the indefiniteness of the superspace metric, or the divergence of functional derivatives in the full theory. {However, we showed by example of a minisuperspace model of flat de Sitter universe that exact normalizable solutions can be easily found and interpreted. In particular, for the de Sitter universe, we found a well-behaved Gaussian solution, which shows a state that bounces back at the time of maximal contraction, thus avoiding the classical asymptotic regression to \emph{nihil}.}

{The next step along this line of research} could be the formalization of our {heuristic} approach both in the Hamiltonian and the Lagrangian formalism, where the wave functional could be constructed in terms of a path integral over foliations of constant multi-fingered time. {On the side of application, it would be fundamental to work out exact solutions that include quantum matter degrees of freedom. Expanding on the de Sitter model introduced here by adding perturbations could be relevant for early inflationary cosmology. Furthermore, beyond cosmology, an application to time evolution during the last stage of gravitational collapse could be of interest.}

\vspace{6pt} 
\funding{{This research received no external funding} 
} 

\conflictsofinterest{The authors declare no conflict of interest.
} 

\begin{adjustwidth}{-\extralength}{0cm}
\reftitle{References}

\end{adjustwidth}

\end{document}